\documentclass[a4paper]{spie}  

\usepackage[]{graphicx}

\title{Extremal noise events, intermittency and Log-Poisson statistics in non-equilibrium aging
of  complex systems}  
 
\author{Paolo Sibani 
\skiplinehalf
Fysisk Institut, Campusvej 55, DK5230 Odense M \\ 
}

\begin{document} 
  \maketitle 

\begin{abstract}
We  discuss    the close link  between
 intermittent events ('quakes') and extremal noise 
fluctuations which has been advocated in  recent numerical and theoretical
work. From the  idea that   record-breaking
noise  fluctuations 
 trigger the quakes,  an  approximate 
analytical  description of non-equilibrium aging  
 as a Poisson process with  logarithmic time
arguments can be derived.  
Theoretical predictions for  measurable statistical properties   of
mesoscopic fluctuations are emphasized, and  supporting
numerical evidence is included  from simulations of short-ranged Ising  spin-glass 
models, of the ROM model of vortex dynamics in type II superconductors, and
of the  Tangled Nature  model of biological evolution.\\ 
{\em To appear in the Proceedings of  the third SPIE International Symposium on
 Fluctuations and Noise, 23-26 May 2005, Austin, Texas }
\end{abstract}


\section{INTRODUCTION}
\label{sect:intro} 

The term \emph{physical aging} was coined nearly three decades ago, 
to describe the  slow drift 
of physical properties of  polymers and  other 
soft materials~\cite{VanTurnhout77,Struik78}. 
The  experimental  spin-glass investigations of 
Lundgren and co-workers~\cite{Lundgren83,Nordblad86,Svedlindh87}, 
and Alba and co-workers~\cite{Alba86,Alba87} spurred further
interest, and aging phenomena  have ever since been 
investigated by  a large 
and growing scientific community.
 Thanks to  these  efforts, it is now widely  realized that aging is a key feature 
of non-equilibrium  dynamics in  complex
systems, including, e.g.\ 
type II superconductors~\cite{Nicodemi01}, 
glasses~\cite{Kob00,Kob00b,Utz00,Berthier02},
granular materials~\cite{Josserand00},
and soft condensed matter~\cite{Cipelletti00,Cipelletti05}. 
The presence of common 
dynamical features across this broad  field 
begs the important question of  
why  microscopically
very different systems should age in  a similar way.

In the following, we  briefly review  theoretical arguments and
 experimental and numerical evidence 
 suggesting   that  
extremal, record-breaking,  fluctuations 
play  a pivotal role in aging   after a deep quench.
The emphasis is on  general ideas, 
and  the original papers\cite{Sibani03,Anderson04,Sibani04a,Oliveira05,Sibani05} 
should be consulted for   further details.

\section{BACKGROUND}
\label{sect:background} 
Complex  systems consist of a large number of degrees of freedom, molecules,
spins, particles etc., which are   coupled  through 
a network of mutual interactions and furthermore interact with an external  source
of white noise, e.g., typically, but not exclusively\cite{Sibani95,Sibani98a,Anderson04}, a heat bath. 
The initial quench, e.g.\ a rapid change of temperature,   is likely to produce 
a state of high frustration and strain,   a state which is unable to locally fulfill 
all constraints imposed by the mutual interactions.
The ensuing \emph{aging} process    tends   to  relax or optimize the 
system,  resulting in a series of   metastable, and gradually more stable  configurations
being visited. Concomitantly, 
 dynamical properties as correlation and response function 
'stiffen', i.e. their rates of change decrease  with the age.
  
Being unable to  re-equilibrate within realistic time scales,
  aging materials  display  a variety of dynamical effects,
which are  largely independent of   microscopic details, but   strongly 
 dependent  on the system age   $t_w$, i.e.\ the time  elapsed from the  initial
quench.  Based on the  relative values of $t_w$ and of the observation time $t$, two 
dynamical  regimes can be identified: 
 For     $t \ll t_w $,    physical averages are  approximately constant,
 and two-time correlations  are related to  the linear response through  
 the  fluctuation-dissipation theorem. 
This 'quasi-equilibrium' regime 
 strongly resembles  the  time translation invariant  dynamics of thermal equilibrium, 
except  that
many  quantities, e.g. relaxation times,  may carry  a parametric age  dependence.
For     $t \gg t_w $,  correlation and response functions 
  no longer   obey the fluctuation dissipation theorem and
  physical averages   drift at a rate which decreases  with increasing $t_w$.

Among the many interesting  facets  of  aging phenomenology, 
the presence of   so-called  \emph{intermittent}
 events in mesoscopic fluctuation spectra\cite{Bissig03,Cipelletti03a,Buisson03,Buisson03a}
 most directly   demonstrates the occurrence 
of  two types of events during the  unperturbed  aging of glassy systems:
 \emph i)  reversible,   quasi-stationary fluctuations 
 near  a metastable configuration, which  
resemble true equilibrium fluctuations and 
 have a Gaussian probability density (PDF),  as in an equilibrium state.  
\emph ii) Jumps, or \emph{quakes} which lead the system, or  
 in a spatially extended system, parts of it,  from one attractor, or 'valley' 
to the next. These  sudden and  large  
configurational re-arrangements  are dominant in the non-equilibrium 
aging regime, and carry the drift 
of macroscopic observables. Their fingerprint  
is  a non-Gaussian, usually exponential, tail of the  
PDF of the mesoscopic fluctuations. 
 
The analysis of  several  specific 
examples~\cite{Sibani93a,Sibani99a,Sibani01,Sibani03,Sibani04a,Anderson04,Sibani05,Oliveira05}   
 has shown the intimate connection between  quakes and extremal, 
record-breaking, noise fluctuation within   metastable
attractors.  Assuming that extremal fluctuations trigger the quakes  
enables one to translate the properties of record statistics into predictions 
of the  statistical properties of the intermittent events,
which are found to be in reasonable  agreement with experimental and numerical
results\cite{Sibani05,Sibani04a,Oliveira05}.

\section{Record-breaking fluctuations and memory in glassy dynamics} 
\begin{figure}
   \begin{center}
   \begin{tabular}{c}
   \includegraphics[height=7cm]{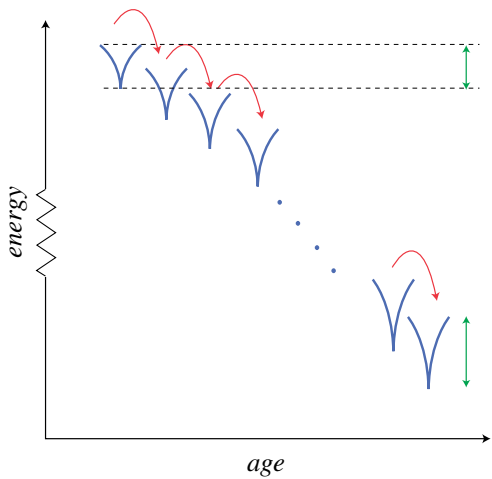}
   \end{tabular}
   \end{center}
   \caption[land_sketch]   
   { \label{fig:land_sketch}An idealized rendering of the
    attractors (wedges) visited by an aging system.  
The arrows represent `quakes'  which lead to the 
new attractors. 
Among a large number of possibilities (not shown)   
 the  marginally more stable attractors are typically selected. 
These quakes  entail a large energy change---the vertical displacement of the wedges---and 
are hence de-facto irreversible. By contrast, the degree of stability, which is
symbolized by the vertical distance between the bottom ad the top 
of each attractor, only increases by an infinitesimal amount.  The layered structure of 
valleys into valleys  likely present in complex systems is 
omitted for typographical simplicity.    
 }
   \end{figure}

How can record-breaking fluctuations have an   impact  
on the time evolution of physical averages? This would e.g. be  plainly impossible
 in thermal  equilibrium, where  averages  do not change. 
  As detailed  below,     two widespread properties 
of  glassy systems, which are sketched  in Fig.~\ref{fig:land_sketch},
 suffice to produce the effect: 
  \emph{i)} the presence of numerous  metastable 'valleys'
  (defined below), 
and \emph{ii)} the   `marginal stability' of these valleys   
against noise fluctuations~\cite{Sibani93a}. 

To fix our  terminology, metastable configurations 
of a noisy dynamical system correspond to
stable configurations  of the noiseless dynamics.  E.g.\ for thermal
hopping,    local energy minima, often called  Inherent
Structures (IS)~\cite{Crisanti00},  are metastable configurations. 
The exit time  associated with the 
basin of attraction of a  metastable configuration  
  i.e.\  the time scale of the
escape probability density,  provides a quantitative
measure of the degree of stability, or depth,
of the basin. In the case of spatially extended systems with 
short-ranged interactions, similar  considerations
  apply to each of several independently 
aging  subsystems,  whose linear size is of the order of the
thermal correlation length. 

To describe the    observed stiffening of   dynamical
properties  in aging systems, it is convenient to
distinguish between the  configurational changes  
leading to  attractors of lesser or equal stability and those leading  to attractors of 
greater stability~\cite{Sibani03}.  Changes of the first kind  are regarded
as  internal rearrangements  of the  `current' valley, while
those of the second type, the quakes, 
 lead  to a new, deeper,  valley.  Clearly, then,   
a configurational change qualifying as a quake at the early 
stages of the dynamical evolution, can at later stages become a reversible fluctuation.  

These concepts  are    translated 
 into a 'non-invasive'  exploration algorithm~\cite{Dall03,Boettcher04}, which
 has uncovered  a highly non-trivial valley structure of spin-glasses,
 and shown that the geometrical information culled from 
time series of energy values   correlates
 well with the results of more traditional  approaches, using e.g.\ 
 repeated quenches\cite{Schon00}.
 Here,   we  note that, by construction,  valleys    grow deeper   with
 increasing $t_w$, and  concomitantly   include more  states. They can
 therefore  be expected to  acquire a layered 
substructure of 'valleys within valleys'~\cite{Dall03,Boettcher04}. 
This  hierarchical organization is  purely a matter of 
 time scales,   as e.g.\   explicitely shown in an example
based on a configurational analysis of the 
Traveling Salesman Problem~\cite{Sibani93}. 

Returning to how record-breaking
fluctuations can leave a permanent mark on the evolution
of   glassy  systems, we  note that condition  \emph{(i)}
ensures that a `sufficiently large' 
fluctuation is able remove the system  
from its current metastable configuration.
The stronger property  \emph{(ii)} means that 
a fluctuation exceeding   all preceding 
fluctuations by an arbitrary  small amount,   hence  a record-breaking 
fluctuation,  is able to  elicit a quake.
  In turn,  this implies that  the \emph{rank}, 
 not the   size, of fluctuations    is crucial for 
aging dynamics. The ability to  retain  a  memory  of 
the largest noise fluctuation occurred during the previous evolution
   can arise through 
a simple entropic mechanism:   As 
the initial quench typically leads to a very  shallow valley,   we can  safely assume 
that the number of valleys 
of depth  $s$  decreases quickly, e.g. exponentially,  with $s$.
The \emph{marginal} increase
 of   the   degree of  stability  of the valley 
  selected in a jump must be then understood in a statistical 
  sense,  and simply reflects    
the overwhelming predominance of  shallow attractors~\cite{Sibani03}. 
In this way the size 
of the largest noise fluctuation experienced so far
 remains imprinted in the height of the  barrier of the current attractor, and the 
 temporal sequence of record-breaking noise events determines
 the overall dynamical evolution\cite{Sibani93a}.  
  
Finally, we  note that the importance of marginally stable attractors for
complex dynamics  was noticed---in a  more restricted 
context---by  Tang et al.\cite{Tang87} in  a  seminal paper  
dealing with a    model of elastically coupled degrees
of freedom, which  was  devised\cite{Coppersmith87} to describe pulse-duration
effects  in charge density wave systems.
 \begin{figure}
   \begin{center}
   \begin{tabular}{c}
   \includegraphics[height=7cm]{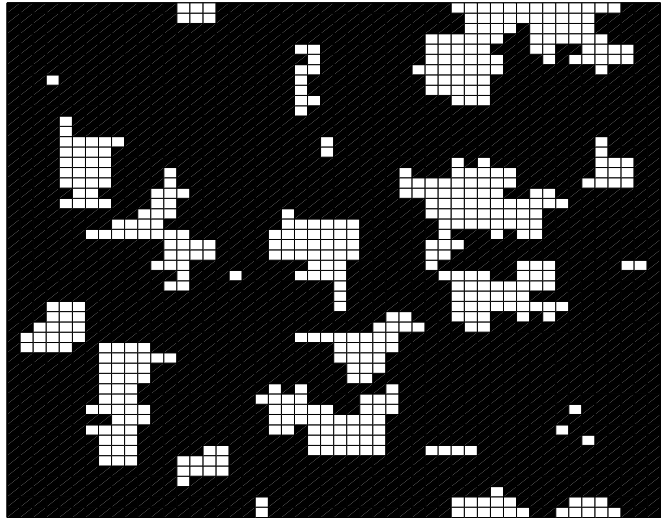}
   \end{tabular}
   \end{center}
   \caption[clusters]   
   { \label{fig:clusters} The
   effect of a quake in a two dimensional 
   Ising spin glass model\cite{Dall03a} is visualized by 
   plotting in white the    overturned spins.   The figure is obtained by    
  projecting   the two lowest-lying states of 
valleys $n$ and $n+1$ onto each other. The data pertain to 
a $2d$ system of size $N=50^2$ at $T=0.3$, with $n=21$.
Note the presence of several  disconnected areas, each comprising
 a relatively large number of spins.
 }
   \end{figure} 
These works initiated a decade of investigations in self-organized criticality\cite{Bak97a,Jensen98},
the study of spatially extended, slowly driven complex systems, which  spontaneously 
organize within  a set of marginally stable states.   SOC   focuses on the power-laws
encountered in the stationary,  scale-free, time evolution of systems whose physical properties
do not change in time, while the log-Poisson theory of aging emphasizes the gradual change
of the stability of the attractors\cite{Sibani01} visited in the aging process.

\section{Record-dynamics as a Poisson process}
In the simple   picture sketched above,  quakes are triggered 
by record-breaking noise fluctuations within each  valley.
A quake entails a large configuration rearrangement, 
which  expectedly   dwarfs   the   
extremal fluctuation which initiates it. Thus, the most obvious manifestation of 
record induced dynamics  lies not in the size distribution 
of the extremal fluctuations,
but in their  temporal distribution.     
Concretely, we seek the 
time dependence of the probability that precisely  
$n$ noise records---and hence quakes---will occur in the 
interval $(t_w,t_w+t]$. 
  \begin{figure}
   \begin{center}
   \begin{tabular}{c}
   \includegraphics[height=7cm]{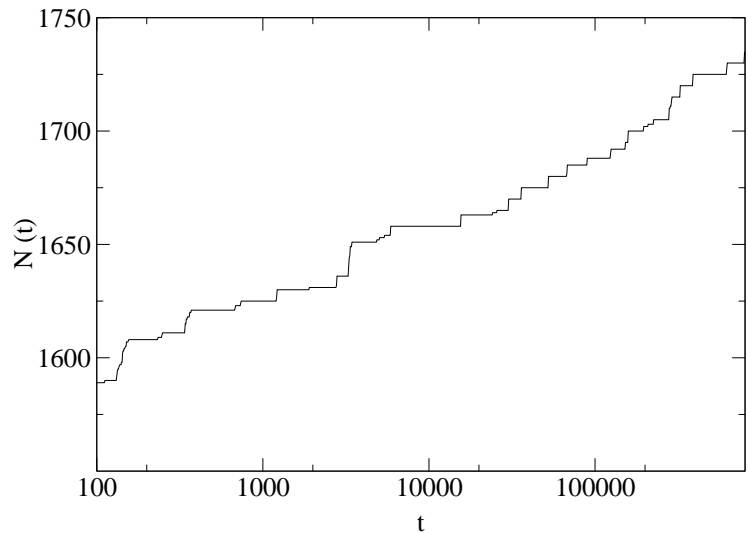}
   \end{tabular}
   \end{center}
   \caption[ROMsteps]   
   { \label{fig:ROM_steps} The number of flux vortices entering 
   the system according to the ROM model of  flux creep in type II superconductors\cite{Oliveira05}.
   This model is chosen to illustrate a generic feature of glassy system:  the 
   quakes  occur in a stepwise fashion. Note the logarithmic scale of the abscissa,
  and the fact that the number of steps per unit of logarithmic time is nearly constant.. 
   The age of the system is $t_w=100$, and the observation time $t$
   exceeds  a million MC steps. We refer to 
   the original article\cite{Oliveira05} for further details.  
 }
   \end{figure} 

To obtain  analytical expressions, 
 the correlation between successive noise events must be neglected.  
This approximation is justified when the noise is small
and quakes are well separated in time. For thermal noise, our
description is thus restricted to a range of  `sufficiently low' temperatures.

With integer times and disregarding correlations,  
 the reversible fluctuations are akin to a  sequence of 
random numbers   drawn  independently   at 
times $1, 2, 3, \ldots, t_w, \ldots t_w+t$ from the \emph{same} distribution.
Crucially, the exact nature of the   noise distribution
from which the numbers are extracted is immaterial~\cite{Sibani93a,Sibani98a}. 
Precisely this mathematical property is at the root of  the 
 generic features of   aging phenomenology.

The probability that $n$ records 
occur in  a sequence of random numbers 
between the $t_w$'th 
to the $t_w +t$'th attempt, 
is  approximately described  by the 'log-Poisson' distribution~\cite{Sibani93a,Sibani98a,Sibani03}  
\begin{equation}
P_n(t_w,t_w+t) =  \frac{(\alpha \log(1 + t/t_w))^n}{n!} (1 + t/t_w)^{-\alpha},
 \quad t>0; \quad  t_w >1.  
\label{logP}
\end{equation}
Strictly speaking, the expression only holds for $n$ small
compared to $t$\cite{Dall03thesis}. We gloss however over this 
restriction, and note that Eq.~\ref{logP} 
  is   a Poisson distribution, whose   average  
\begin{equation}
 \langle n(t_w,t)\rangle =  \alpha \log(1 + t/t_w) 
\label{average1}
\end{equation}
depends on the difference of the logarithms of the endpoints
of the observation interval $(t_w,t_w+t]$, rather than  
on the linear difference $t$  usually encountered.  
The  free parameter  $\alpha$ is  interpreted as follows:
The number of records in 
$p =1,2 \ldots$   sequences updated independently and in  parallel  is  
itself a  log-Poisson process 
with  $\alpha =p$, and the   value $\alpha  >1$  can cover  spatially extended systems, 
where several regions can evolve  in parallel.
To include the possibility that not every  noise record triggers a quake,
 $\alpha$ is  also allowed to assume non-integer values.
  
\section{Measurable properties}
The relevance of record-breaking 
for aging dynamics  has been  verified  
numerically for several  models~\cite{Sibani93a,Sibani01,Sibani99a,Anderson04,Oliveira05}. 
Below, we mainly   discuss   
predictions which  are testable  by a  statistical analysis  
of mesoscopic fluctuations.  
It is assumed  that relevant 
  time and age dependent statistical fluctuation properties of a  glassy system
 can be  obtained  through  an   ensemble of independent measurements, and that 
these measurement yield a PDF  with a 'normal',  
  typically Gaussian part, which stems from the reversible
 fluctuations, and an 'anomalous', intermittent tail   solely due to
 non-equilibrium intermittent events. Exploiting the large
difference in typical size between the two types of events, 
each data point is classified as either a fluctuation or 
a quake, and the  number of 'Gaussian'
and intermittent events observed  are denoted by $n_G$ and $n_I$ respectively. 

The shape of the intermittent tail  and the
relative weight of the two parts  clearly depend
on the number of quakes falling within the observation   interval.
 This influence is exclusively expressed through the  average
 $\langle n (t_w,\delta  t)\rangle$, the only 
  parameter entering  Eq.~\ref{logP}. 
   \begin{figure}
   \begin{center}
   \begin{tabular}{c}
   \includegraphics[height=7cm]{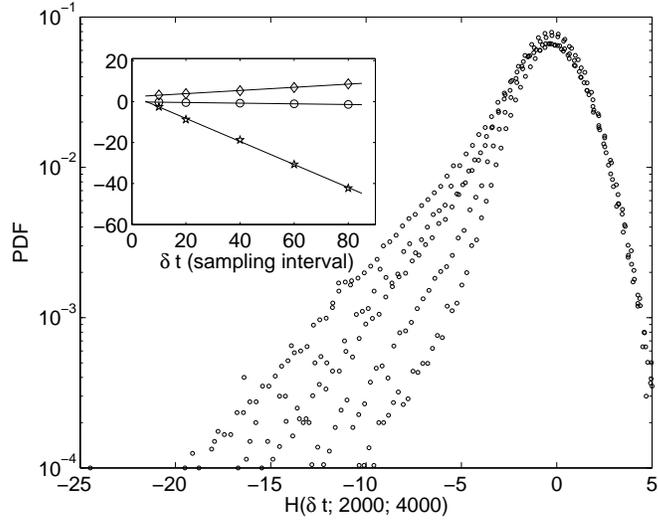}
   \end{tabular}
   \end{center}
   \caption[raw-data]   
   { \label{fig:non_collapse} The data plotted  describe the distribution of 
   the  heat exchanged in spin-glass model~\cite{Sibani05} over a small 
   (dimensionless) time interval $\delta t$, ranging through the
   values $10, 20, 40, 60 $ and $80$ 
   for the five data set shown.
    The system has   age $t_w = 4000$,
   and the statistics is collected over a time interval $t=2000$. 
   The insert show the three central moments, average variance and skewness, 
   of the distribution, as a function of $\delta t$.  
}
   \end{figure} 

   \begin{figure}
   \begin{center}
   \begin{tabular}{c}
   \includegraphics[height=8cm]{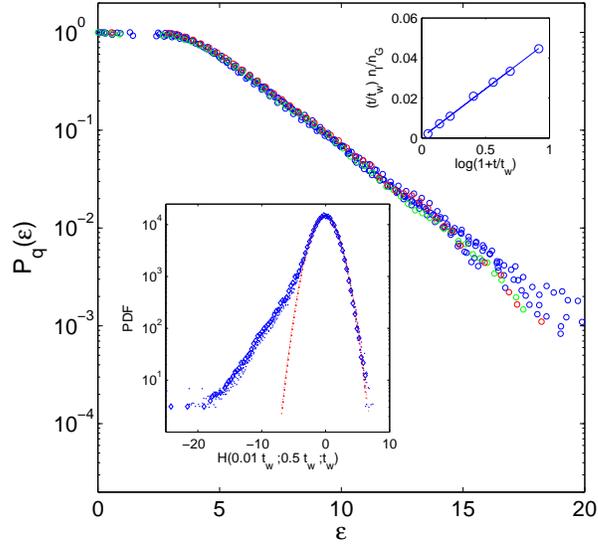}
   \end{tabular}
   \end{center}
   \caption[collapse]   
   { \label{fig:collapse}  All data plotted describe properties of the 
   PDF of the  heat exchange in a spin-glass model at temperature $T=0.3$\cite{Sibani05}.
   Main panel, and lower insert:  The data collapse shows that the statistical 
   properties of the heat exchanged in a system of age $t_w$ over a time interval
   $\delta t \ll t_w$ only depend on the ratio $\delta t/t_w$. In the main panel,  
   the probability of energy $\epsilon$ being released in a single intermittent
   event is shown to decay exponentially with $\epsilon$. These data are extracted from
   the tail of the distribution. The upper insert is a numerical verifications of Eq.\ref{ratio}.
   The circles are data points, which fall on a fitted straight 
   line, as predicted by the theory.  
 }  \end{figure} 
We first consider a sufficiently small   observation
interval $\delta t << t_w$ , i.e.\ such that 
  the $\delta t$ dependence of  $\langle n (t_w,\delta  t)\rangle$  
can be linearized   as
\begin{equation}
 \langle n (t_w,\delta  t)\rangle =  \frac{\alpha \delta t}{t_w}.  
\label{average2}
\end{equation} 
Accordingly,    PDF data obtained for different 
observation times $\delta t$ and different ages $t_w$   
collapse, whenever  the 
ratio $\delta t/t_w$ remains  constant. This  effect is demonstrated in the lower
insert of   Fig.~\ref{fig:collapse}, which displays the PDF of the heat transfer in 
the three dimensional Edwards-Anderson short-ranged  
spin-glass model~\cite{Sibani05} over a time  interval $\delta t$ for several 
pairs of values of $\delta t$ and $t_w$ with $\delta t/t_w = 0.01$.

Secondly, we consider varying   the observation time  $t$, while keeping  
 $t_w$ at a fixed value.  The number of intermittent events
observed grows on average  as $n_I \propto  \log(1 + t/t_w)$.  
Reversible fluctuation  occur   at a constant
rate, and their number grows as  $n_G \propto t$.   
Hence, the  ratio $ n_I/n_G$, which is, basically, the ratio
between the statistical weight of the tail and the bulk of the
PDF,   is  given by
 \begin{equation}
   \frac{n_I}{n_G} \frac{t}{t_w}  \propto   \log(1 + t/t_w).
 \label{ratio}
 \end{equation}
The plots in the  upper insert of    Fig.~\ref{fig:collapse}
confirm the linear relationship predicted by Eq.\ref{ratio}.
The circles  are   simulation data, which are 
 are  plotted versus 
$\log(1 + t/t_w)$, together with a linear fit (the full line).
 The main panel in the figure shows that the energy $\epsilon$ 
which is liberated in a single quake
has an exponential distribution. 
   \begin{figure}
   \begin{center}
   \begin{tabular}{c}
   \includegraphics[height=7cm]{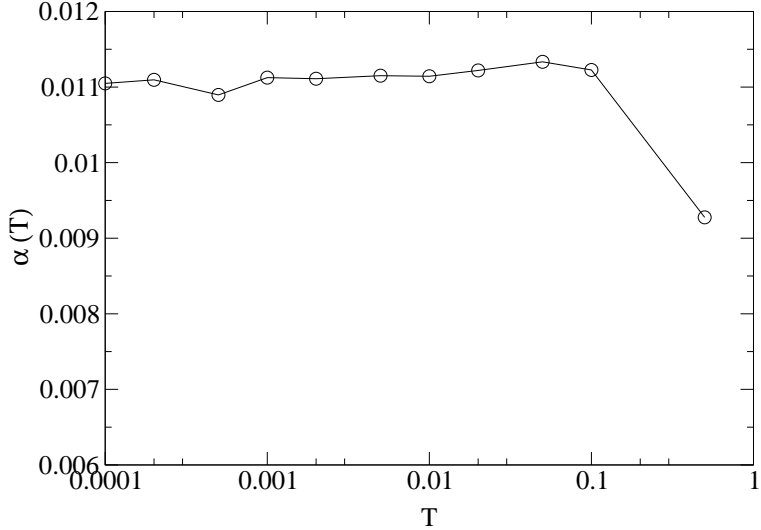}
   \end{tabular}
   \end{center}
   \caption[ROM]   
   { \label{fig:ROM_rate} The rate of quakes $\alpha(T)$ as a function of temperature for
   the ROM model of flux creep in type II superconductors. The  rate is 
   seen to be nearly temperature independent over a wide range 
   of temperatures. The plot is taken from Oliveira et al.\cite{Oliveira05} 
 }
   \end{figure} 	  
The above  properties  are all   independent
of the temperature, as  e.g. \  observed in 
 short-ranged spin    glass models\cite{Dall03,Sibani05,Sibani04a}. 
 To illustrate the independence,  we show  the rate at which 
 flux  vortices enter the system  in  the 
 Random Occupancy Model (ROM),    a thermally activated model 
 of  vortex dynamics in type II superconductors\cite{Nicodemi01,Jensen02}
 which  reproduces the essential experimental 
 features of these systems.
 A detailed description is given
 by  Oliveira et al.\cite{Oliveira05}, from which  Fig.~\ref{fig:ROM_steps} and
  Fig.~\ref{fig:ROM_rate}  are borrowed. 
 The  vortex creep   occurs in the  stepwise fashion 
 shown in Fig.\ref{fig:ROM_steps}, where  each  step corresponds to a quake leading 
 to a new metastable configuration.  Fig.~\ref{fig:ROM_rate} shows  that the
 rate at which quakes occur  remains constant over  a wide range of 
 temperatures. 
 
An activated process with temperature independent rates 
sounds at first  as an oxymoron, but a second look 
gives   insight into  the  subtle role 
played by the local density of states (LDOS) within each metastable attractor,
and, more generally, into  the memory behavior of complex systems:
Importantly,  record dynamics  only   claims   $T$ independence
for  the rate at which \emph{extremal} barriers are climbed. Unlike generic 
barriers, these barriers are determined by the noise history, 
and  hence grow   with $t_w$  in a temperature
 \emph{dependent} fashion. As the    temperature increases,
 this barrier growth and 
 the decrease of the  Arrhenius time scale for barrier 
 crossing must cancel  
 in order to produce the  desired effect. Specifically, to get a 
 cancellation,  the  LDOS for
all the attractors   must have  the exponential form\cite{Sibani05}  
 \begin{equation}
      {\cal D}(\epsilon)\propto \exp(\epsilon/\epsilon_0). 
      \label{expD}
      \end{equation} 
Note that  thermal metastability is then restricted
to the temperature range     $T <\epsilon_0 $.  

The  Boltzmann  distribution
	for  an  exponential LDOS is given by 
     $P_E(b) = \exp(-b a(T))$, where         
     $a(T) = (1/T -1/\epsilon_0) > 0$ is an 
      effective inverse temperature.
     Secondly,  with a constant microscopic attempt rate, 
      the largest   fluctuation 
     on a  time scale $t_w$ is  the largest   
     of  ${\cal O}(t_w)$ values drawn independently from  
        $P_E$. This extremal fluctuation is identical in size
	to the extremal barrier characterizing the attractors visited
	at age $t_w$. 
    Applying a  well known mathematical result~\cite{Leadbetter83},
    the typical extremal barrier climbed at time $t_w$  has height 
    \begin{equation}
     b_{max}(T,t_w) \propto \ln(t_w)/ a(T).
     \label{extremal}
     \end{equation} 
   The corresponding Arrhenius time is therefore 
    \begin{equation}
     \tau_A  \propto \exp(a(T) b_{max}) = t_w, 
     \label{extremal_time}
     \end{equation} 
   as anticipated. 
\begin{figure}
   \begin{center}
   \begin{tabular}{c}
   \includegraphics[height=7cm]{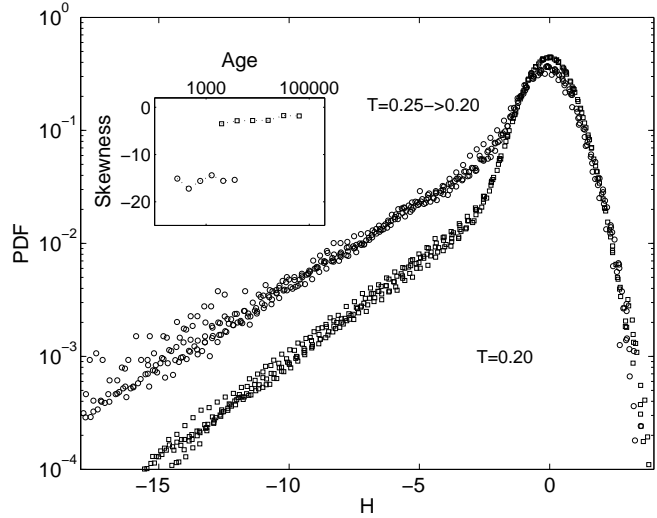}
   \end{tabular}
   \end{center}
   \caption[effectiveage]   
   { \label{fig:effectiveage} The figure illustrates 
   the effect of a small negative temperature shift (main panel, upper data set) 
   imposed on an aging spin  glass model\cite{Sibani04a}
   and contrasts it to isothermal aging (lower data set). 
   All PDF describe the  heat transfer over   short time $\delta t$,
   $H(\delta t, t, t_w)  = E(t_w+(n+1) \delta t) -E(t_w+n \delta t)$,
$n=0,1,2 \ldots$, where $E$ is the total energy.
   Main figure, lower data sets:   
 The five sets of data shown  correspond to ages
$t_w=100 \delta t$ with 
$\delta t = 10,20,40 \ldots 160$  and 
observation window $t=t_w/2$. The temperature is $T=0.20$.
Upper data sets:   
One additional  cooling step  takes the system from   $T=0.25$ to 
$T=0.20$.  For the $i$'th data set, $i=1,2, \dots 5$ 
the negative shift occurs at age  $t_{wi}$, chosen  such that 
  the corresponding 
\emph{effective} ages obtained according to Eq.~\ref{effective_age}  are 
   $t_w^{eff}= 2000, 4000, 8000, 16000$ and  $32000$. The
corresponding sampling intervals are 
$\delta t =t_w^{eff}/100$ and $t=t_w^{eff}/2$.  
Keeping the ratio  $\delta t/t_w^{eff}$ constant collapses the  data,
whence the $t_w^{eff}$ plays the same role as the
$t_w$ does in the isothermal case.
The insert shows that the skewness of the PDF's 
 falls into two distinct groups corresponding to $T$ shifted
 and isothermal aging, with only a small variation
within each group.
 }
\end{figure} 

Interesting effects arise when an aging system is perturbed by
small variations of external parameters as e.g. the temperature\cite{Takayama02,Sibani04a}.     
 A small negative temperature  shift to $T'<T$  imposed
 at $t_w$  does not change the  extremal barrier 
      already established, but  affects   the spectrum of thermal fluctuations
        available to cross it\cite{Sibani04a}. 
	We   immediately find 
 \begin{equation}
     \tau_A  \propto \exp(a(T') b_{max}) = t_w^x, 
     \label{effective_age}
     \end{equation} 	
where the exponent $x = \frac{a(T')}{a(T)}>1$   defines the effective age
\begin{equation}
     t_w^{eff} = t_w^x, 
     \label{effective age }
     \end{equation} 	
 The above argument  does not cover a  
  positive  temperature shift $T' > T$. In this case, the ensuing   stronger  fluctuations  
quickly   overcome the extremal barrier  $b(t_w,T)$ previously
established,   and reach  the extremal barrier $b(t_w,T')$
for  isothermal relaxation  at the final temperature $T'$\cite{Sibani04a}.

 An approximately  exponential LDOS  is    found in numerical investigations
  of various complex systems\cite{Sibani93,Sibani94,Sibani98,Klotz98a,Schon00}.
 Secondly, since the LDOS determines the quasi-equilibrium fluctuation spectrum, it
 is in  principle available from calorimetry measurements, which for a spin-glass
 system lead  to the same conclusion\cite{Sibani05}.
  Thirdly,  since a quake is only irreversible
  on the time scale at which it occurs, the same 
  process  can occur  reversibly within an enlarged valley at a later stage. Hence,
  the exponential distribution of  the energy released  in a quake, see e.g. Fig.\ref{fig:collapse},
    is  itself concurring, albeit indirect,   evidence for the ubiquity of exponential density 
  of states in complex systems.
      
\begin{figure}
   \begin{center}
   \begin{tabular}{c}
   \includegraphics[height=7cm]{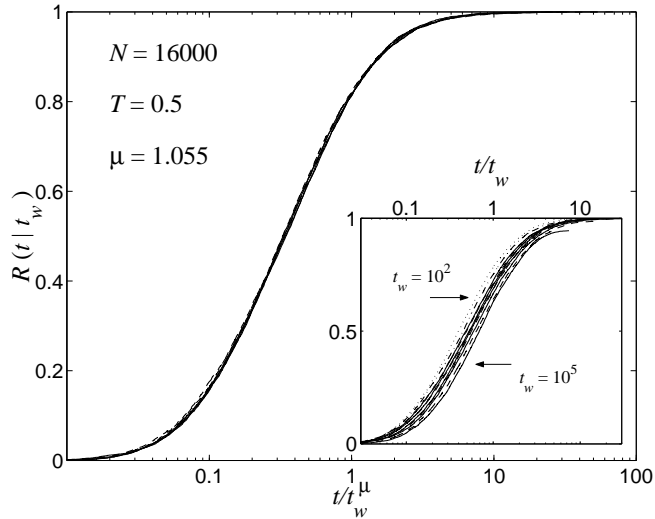}
   \end{tabular}
   \end{center}
   \caption[residence]   
   { \label{fig:residence} The 
   empirical distribution of the residence time in a valley $t_r$, 
   plotted vs. the scaled variable $t/t_w^\mu$ in the main
   panel and vs.  $t/t_w$  in the insert. A wide range of $t_w$ values
   was considered. The system in question is a spin    glass on a $k$-regular random graph, 
   but the same type of results is found for Euclidean lattices in $2$ and $3$ spatial dimensions.
   }
   \end{figure}

 Let us now consider the statistics of the time intervals between intermittent 
 events. Its precise  determination  requires the ability 
 to discriminate between fluctuations and quakes, and improved experimental techniques
 are required to do so\cite{Buisson04}. 
 Within a   log-Poisson theoretical description, the theoretical   
 probability to find an  interval $t_r$ between   consecutive quakes$/$intermittent events
 which is  smaller  than $t$ 
 is given by\cite{Sibani03,Dall03} 
 \begin{equation}
 R(t \mid t_w) = 1 -( 1 + \frac{t}{t_w})^{-\alpha}
 \label{an_res}
 \end{equation}   
where $\alpha >1 $ was  introduced in Eq.~\ref{logP}.
Note that the average time between the events is $\langle t_r \rangle = t_w/(\alpha -1)$,
which is finite  for $\alpha >1$. Theoretically, the  distribution depends on 
the ratio $t/t_w$ only, which is  so-called \emph{pure aging} behavior.
Equation~\ref{an_res}  has been checked  
numerically\cite{Dall03} with a typical result 
plotted in  Fig.\ref{fig:residence}: The insert shows that a 
reasonable  collapse of the empirical data can be obtained using  $t/t_w$ scaling. 
At this level of approximation, the  analytical expression  cannot be distinguished
 from the data.   
 The    better collapse shown in the main panel is obtained
 with the scaling variable $t/t_w^\mu$, where $\mu$ is a weakly size dependent parameter
 close to unity. For the chosen parameters,
 $\mu = 1.055$. Other numerical simulations\cite{Dall03b} indicate that 
 the small deviations from $t/t_w$ scaling 
 decrease systematically as the system size grows and/or the temperature decreases.
Experimental deviations are observed in 
several other cases\cite{Cipelletti05}. In spin-glasses they 
are attributed to the finiteness
of the cooling rate\cite{Rodriguez02,Zotev03}. Such effects   
are not covered by this theory in its present form. 
   Note that   Figs.\ref{fig:clusters} and \ref{fig:residence}
are not obtained by an intermittency analysis, but stem from the 
numerical exploration algorithm\cite{Dall03} previously
mentioned. The close relation between the 
two approaches is briefly discussed below.

\begin{figure}
   \begin{center}
   \begin{tabular}{c}
   \includegraphics[height=7cm]{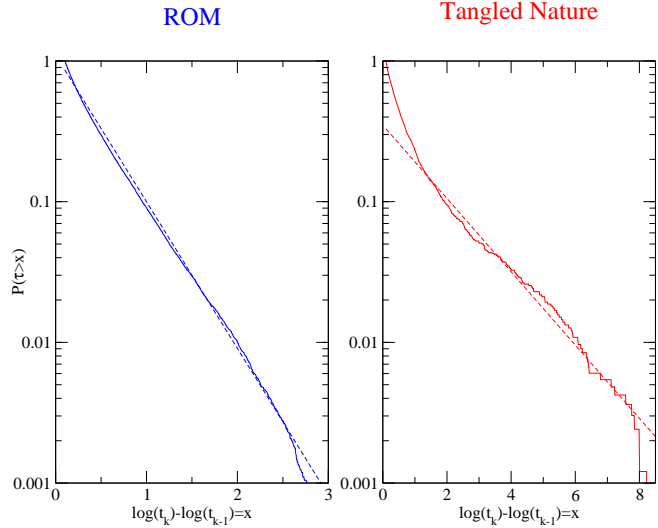}
   \end{tabular}
   \end{center}
   \caption[logwt]   
   { \label{fig:logwt} 
The empirical distribution for the  log waiting times numerically obtained for  
 the ROM and the Tangled Nature model\cite{Hall02,Anderson04} is shown.
In good agreement with Eq.\ref{lw}, the results  nearly follow an exponential 
decay (broken line).  
 }
   \end{figure}

As further evidence of the relevance of the Log-Poisson description for complex dynamics,
 we finally consider the distribution of the 'log waiting times', which are defined as follows:
 Let $t_1 \ldots t_k$ be the times at which quakes $1 \dots k$ occur.
  The waiting times  between the
 events are independent and  exponentially distributed stochastic variables 
 in a usual Poisson process, while 
 in  a Log-Poisson  process the same  is true for the 'logarithmic waiting times' defined as
 \begin{equation}
 \tau_k = \log(t_k) - \log(t_{k-1}) = \log(t_k/t_{k-1}).
 \label{lw}
 \end{equation}
 In principle, this distribution is   measurable  in intermittency experiments   
 by recording the  intervals between anomalous spikes.
Numerical results for  the ROM model previously described and for
the Tangled Nature model of biological evolution\cite{Hall02} are shown 
in Fig.\ref{fig:logwt}. 

\section{Summary  and Conclusion}  
The  Log-Poisson  non-equilibrium aging scenario  leads, in a general 
and near parameter-free manner, to   analytical 
expressions for  several  statistical properties of mesoscopic fluctuations 
which are of relevance for the intermittent behavior of glassy systems
after a deep quench. The  distinction between time homogeneous  fluctuations and time heterogeneous quakes 
is built right into  the description, which hence  naturally  covers
various  types of complex dynamics, where the same distinction applies, 
  e.g.\   driven dissipative systems\cite{Sibani93a,Sibani01},
and biological evolution\cite{Sibani95,Sibani99a,Newman99c,Anderson04,Anderson04thesis}.  
Broadly speaking, the physical mechanism behind all these 
 examples combines an entropic aspect,  the over-abundance
 of  attractors with low stability, and a   dissipation (of e.g.\ energy)  mechanism,
 which guarantees  the ability of the system to spontaneously 
  drift away from its initial  state.  
   These  aspects  have not received  much attention in   other  theories of  aging,   where
  the main  focus is on the linear response to a 
  small   external perturbation\cite{Cugliandolo97,Calabrese04}.
 In spite of this important difference, the   log-Poisson  description also  adapts and$/$or incorporates   
 other  established  ideas about  complex dynamics.
 The emphasis on marginal stability modifies and extends pre-SOC developments\cite{Tang87,Coppersmith87}
 on memory behavior.   The increasing level of 
 hierarchical organization acquired by the valleys  is  a recurring theme in complex
 dynamics\cite{Sibani89,Joh96}. Finally, attractors (and valleys) 
  are defined at the level of thermally 
 correlated clusters\cite{Sibani05}. These are  localized  objects separated by  
 frozen degrees of freedom, and their increasing stability arguably corresponds to  the
 growth of the thermal correlation length\cite{Berthier02}. 
 
 Several questions are under investigation, e.g. more detailed predictions for
 the fluctuation spectrum of correlation estimators used in measurements\cite{Cipelletti05},   
 the origin of the exponential density of states and extending the picture to the case
 where attractor properties beyond the stability can change as the system ages.
 
\acknowledgments     
P.S. thanks  Jesper Dall, Henrik Jeldtoft Jensen, Paul Anderson,  Luis Oliveira and Stefan Boettcher
for their collaboration on many aspects of complex dynamics. This work was supported by a  grant  from
the Danish SNF. 

\bibliography{SD-meld,thesis}   
\bibliographystyle{spiebib}   

\end{document}